\documentclass[journal]{IEEEtran}
\usepackage{xcolor}
\usepackage{amsmath}
\usepackage{algorithm}
\usepackage[noend]{algpseudocode}
\usepackage{afterpage}
\usepackage{multirow} 
\usepackage{url}
\usepackage{cite}

\usepackage[justification=centering]{caption}

\ifCLASSINFOpdf
  \usepackage[pdftex]{graphicx}
\else
  \usepackage[dvips]{graphicx}
\fi

\usepackage{amsmath}
\usepackage{array}

\usepackage[export]{adjustbox}

\usepackage[caption=false,font=footnotesize]{subfig}

\usepackage{float}

\begin{document}

\title{Improving BLE Beacon Proximity Estimation Accuracy through Bayesian Filtering}

\author{Andrew~Mackey,~\IEEEmembership{Student Member,~IEEE}, Petros~Spachos,~\IEEEmembership{Senior Member,~IEEE}, Liang~Song,~\IEEEmembership{Member, IEEE}, and Konstantinos N. Plataniotis,~\IEEEmembership{Fellow,~IEEE} 
                
\thanks{This work was supported in part by the Natural Sciences and Engineering Research Council (NSERC) of Canada.
\par A. Mackey and P. Spachos are with the School of Engineering, University of Guelph, Guelph, ON, N1G2W1, Canada.
(e-mail: mackeya@uoguelph.ca; petros@uoguelph.ca).
\par L. Song is with the Fudan University, Shanghai, China. (e-mail: lsong@ieee.org)
\par K. Plataniotis is with the Department of Electrical and Computer Engineering, University of Toronto, Toronto, ON, M5S3G4, Canada. (e-mail: kostas@ece.utoronto.ca)}}

\maketitle

\begin{abstract}
The interconnectedness of all things is continuously expanding which has allowed every individual to increase their level of interaction with their surroundings. Internet of Things (IoT) devices are used in a plethora of context-aware application such as Proximity-Based Services (PBS), and Location-Based Services (LBS). For these systems to perform, it is essential to have reliable hardware and predict a user's position in the area with high accuracy in order to differentiate between individuals in a small area.  A variety of wireless solutions that utilize Received Signal Strength Indicators (RSSI) have been proposed to provide PBS and LBS for indoor environments, though each solution presents its own drawbacks. In this work, Bluetooth Low Energy (BLE) beacons are examined in terms of their accuracy in proximity estimation. Specifically, a mobile application is developed along with three Bayesian filtering  techniques to improve the BLE beacon proximity estimation accuracy. This includes a Kalman filter, a particle filter, and a Non-parametric Information (NI) filter. Since the RSSI is heavily influenced by the environment, experiments were conducted to examine the performance of beacons from three popular vendors in two different environments. The error is compared in terms of Mean Absolute Error (MAE) and Root Mean Squared Error (RMSE). According to the experimental results, Bayesian filters can improve proximity estimation accuracy up to 30\% in comparison with traditional filtering, when the beacon and the receiver are within 3 m.
\end{abstract}


\IEEEpeerreviewmaketitle

\section{Introduction}
\IEEEPARstart{W}{ith} the current advancements in Wireless Sensor Networks (WSNs), buildings, homes, and many other indoor environments have become a new platform to which electronic devices and occupants can share data and location information to improve things ranging from customer experience to individual quality of life and satisfaction~\cite{kortuem, jin}. This may include targeted advertising in the commercial space or individualized lighting and temperature control in the home, based on room occupants. These services that can be provided in an indoor environment rely on the location of the individual(s). Unfortunately, the Global Positioning System (GPS) does not work in an indoor environment due to physical obstruction, nor does it meet the necessary sub-meter accuracy requirements. For scalability, an alternative solution that is accurate, simple, cheap, and power-efficient must be developed. Deployment of such systems must be feasible in buildings ranging from small homes to large malls and industrial-sized buildings~\cite{ xu}. This leads to the emergence of smart buildings and smart homes, all falling under the realm of the Internet of Things (IoT)~\cite{spachos, zanella, naphade, alfuqaha}. 
 
There are indoor location solutions for GPS-less environments based on wireless technologies such as Wi-Fi and vision/ camera dependent systems~\cite{liu, davidson, RGB_D_SLAM,singleCamera,VisualVocabLoc,wifiLoc}. However, these systems have drawbacks, such as the higher hardware costs associated with cameras~\cite{RGB_D_SLAM}. The requirement of pre-computed environmental data associated with vision-based systems~\cite{singleCamera}, or the high power consumption inherent in Wi-Fi-based solutions~\cite{EnergyCompare}. 

In recent years, Bluetooth Low Energy (BLE) beacons, commonly known as beacons, increase their popularity due to their high availability, low cost, low power consumption, and ease of deployment~\cite{spachos}. Beacons can operate on coin cell batteries or even batteryless~\cite{spachossolar, jeon}.
Many context-aware applications are using beacons, such as Proximity-Based Services (PBS)~\cite{pai, Ghose} and Location-Based Services (LBS)~\cite{alletto,faragher, xiao, he, particleFilt,trackingApp, GuassianKalman, NN_knB,3D}, especially in GPS-less environments. Both services might use similar infrastructures, like beacons, however, they have different underlying operational logic.  LBS  deliver information according to the location of the user. On the other hand, PBS deliver information according to the proximity of the receiver node from the transmitter node. 

  
This paper examines the usage of three Bayesian filters, namely, the Kalman filter, the particle filter, and the Nonparametric Information (NI) filter, implemented on a mobile application, to improve the accuracy of BLE beacons for PBS in GPS-less environments. All experiments are conducted with the use of three beacon hardware devices that implement Apple's iBeacon protocol: Estimote~\cite{estimote}, Kontakt~\cite{kontakt}, and Gimbal~\cite{gimbal}, all configured with the same transmit power and transmit interval, -12 dBm and 100 ms, respectively. The receiving device should be Bluetooth 4.0 compatible, hence, a Google Nexus~5 smartphone is used. An Android application is developed to measure the RSSI values and convert them into observable distance estimations. The RSSI values are passed into the filters in order to better estimate user proximity. All application functions and filters are developed and executed in the developed mobile application. Experimental results show an improvement in proximity accuracy up to 30\%, over the estimations based on simple moving averages when the beacon and the receiver are within 3 m.

The novelty of this work relies on the  experimental comparison of the beacons along with the usage of the Bayesian filtering techniques. The rest of this paper is organized as follows: In Section~\ref{related}, an overview of the related work is given. The problem formulation is in Section~\ref{problem}, followed by Section~\ref{filtering} with a brief description of the filtering techniques. Section~\ref{system} presents the experimental setup  along and a discussion and analysis are in Section~\ref{exp}. The conclusion is in Section~\ref{concl}.

\section{Related Work}\label{related}

An extensive overview of the current technologies, techniques, and services associated with IoT equipped smart buildings is given in~\cite{spachos, smartBuildings}. Applications regarding tenant services, disaster management, marketing, and personal security are discussed to provide a basic understanding of the current issues and potential solutions in this field. Beacons are wireless devices receiving a lot of attention for IoT applications, capable of providing solutions for a lot of the aforementioned IoT applications in smart-buildings~\cite{ng}.  The survey presented in~\cite{kang} examines deployment cases, hardware requirements, and distance estimation methods of beacon devices. It introduces localization, proximity detection, and activity sensing as the primary application uses for beacons.

In~\cite{ramsey},  the development and testing of location fingerprinting using BLE beacons is shown,  where fingerprinting relies on a previously constructed radio map of the deployment region. The authors use 19~beacons in a 600~$m^2$ environment. The best results showed a positioning error under 2.6~m in a dense beacon network. In~\cite{pai}, the authors utilize beacons to design a proximity detection system. They stress the effects of interference between beacons in a dense network. Experimental results show an achievable detection accuracy of 90\% under the employment of their adaptive scanning mechanism. The adaptive scanning accounts for changes in beacon density, so as not to enforce a constant scan interval, ensuring observations are made during the scan. 

A different approach combines technologies with beacons in order to provide more accurate positioning. In~\cite{rohan}, the authors use inertial sensors with BLE beacons in order to mitigate signal nonlinearities and drift error. A Trusted K-Nearest Bayesian Estimator is used to estimate position. This method first obtains the position using the K-Nearest Neighbor method and then applies a fusion of the pedestrian dead-reckoning approach and Kalman filter. The positioning error was found to be under 1 m.

Proximity accuracy is often improved through the use of filters. In~\cite{particleFilt}, a particle filter is implemented to enhance the RSSI accuracy of an iBeacon based micro-location system. By experimentation using Gimbal 10 series beacons, the optimal particle, and beacon number selection are determined, thus improving the overall position accuracy. For real-world applications, the importance of beacon placement at high altitudes to avoid obstacle interference was stressed. Additionally, an Android application with real-time remote tracking capabilities that implements a Kalman filter for improved accuracy has been developed. The experiments present an attainable accuracy in the order of meters~\cite{trackingApp}. Similarly, in~\cite{GuassianKalman}, three popular filtering techniques: Kalman filter, Gaussian filter, and a hybrid of the two are directly compared. The experiment was comprised of a Texas Instruments CC 2540 iBeacon base station used in three separate environments. The Kalman-Gaussian Linear filter proved to be the most effective at reducing system noise, ergo improving indoor location accuracy. In~\cite{lam}, an optimized support vector machine (O-SVM) on the cloud for distance estimation and a Kalman filter (KF) on the edge to obtain a near true RSS value from a list of RSS measurements is proposed. Additional works experiment with the raw accuracy of BLE RSSI signals for location purposes and suggest implementing a simple smoothing algorithm to obtain better accuracy~\cite{smoothing, spachosRSSI, spachosicassp}.

In this paper, we improve upon the accuracy of BLE beacon-based indoor proximity estimation by implementing  three Bayesian filtering techniques. The results of each filter implementation are compared in order to find the best suitor for an accurate, simple, and easy to deploy beacon-based PBS. 

\section{Problem Formulation}\label{problem}
Proximity estimation solutions are often hard to scale and the distribution of such systems needs to be robust, precise, and fitting to all environments. Utilizing RSSI techniques is difficult due to nonlinearities in the system. These nonlinearities can be a result of LOS obstruction, signal interference/ noise, and dynamic environmental conditions. These dynamic properties are the major governing factors in the performance of RSSI-based estimation. As such, the RSSI of the wireless signals broadcast by beacons follow the subsequent path loss model~\cite{gu}:

\begin{equation}
RSSI = - 10\ n\ log_{10}(\frac{D}{D_{0}})+ C_{0} 
\end{equation}
where $RSSI$ is the observed RSSI in dB, $n$ is the path-loss  exponent that corresponds to the environment, $D$ is the distance between the beacon and the user, $D_0$ is the reference distance and $C_0$ is the average RSSI value at the reference distance.

The first problem in modeling the environment is to optimally define the $n$ and $C_0$ parameters of the path-loss model through experimentation with each beacon. Then, at a reference distance, $D_0$, of 1 m, proximity and distance estimation can be calculated as: 

\begin{equation}
D = 10^\frac{C_0-RSSI}{10\ n}\label{equation}
\end{equation}

The second problem is obtaining accurate distance estimations $D$ by ensuring the observed RSSI  is ideal. State estimation filtering techniques are applied in order to mitigate the effects of nonlinearities in the system, thus, resulting in better proximity estimation.

%
%

\section{Filtering Techniques}\label{filtering}
This section describes the filtering techniques applied to the observed set of RSSI values in all experiments.

\subsection{Simple Moving Average}
The simplest and most common filtering algorithm that can be implemented is a simple moving average (SMA). This is the running average of the last $N$ measurements, where $N$ is the pre-defined width of the SMA. It is given as:

\begin{equation}
RSSI_{SMA} = \frac{\sum_{i=1}^NRSSI_i}{N}
\end{equation}

SMA was implemented to be used as a base case and a simple filtering technique, in comparison with the other filters. Through experimentation, it was found that a window size of 20 measurements is ideal for our experiments.

\subsection{Kalman filter}\label{kalman}
Kalman filter is a recursive stochastic filter optimally suited for posterior state estimation of linear and Gaussian systems. It is a continuous Bayesian estimator that determines the optimal Minimum Mean Square Error (MMSE) estimate of a multivariate system. In our case, it can estimate a sequence of proximities over time, $p(1),p(2),...,p(k)$, given a sequence of noisy RSSI values, as:

\begin{equation}
R(k) = \{r(0),r(1),...,r(k)\}
\end{equation}

The estimated proximity $p(k)$ is conditional on the observations. Thus, $p(k|j)$ is the estimated proximity at time $k$ given the record of all RSSI observations leading up to and including $R(j)$. Although RSSI is a nonparametric variable due to path loss, the Kalman filter's straightforward computation and tunable parameters for measurement and process noise make it useful for proximity estimation using RSSI techniques in a noisy environment. 

The algorithm incurs low computational load, making it an ideal suitor for a fully mobile-based implementation. Previous research has shown the implementation of Kalman filters on a server~\cite{particleFilt}. However, in our work in this paper, the Kalman filter is implemented in the mobile application. The mobile application is designed to measure the RSSI values transmitted by the beacons and translate them into distance estimations. Most of the smartphones on the market are equipped with the necessary connectivity (i.e. Bluetooth capable antennas, and Bluetooth 4.0) to act as the receiver. Furthermore, in utilizing smartphones we eliminate the need for additional hardware so that it is easily implemented and scaled for all potential users of such indoor location systems.

Kalman filter has two stages, prediction and update (correction). The filter first makes a prediction at time $k$ of the current state $x$ and the system covariance $P$ based on the prior before a new RSSI observation $r(k)$ is made. The second stage computes a gain value $G(k)$ based on the prior noise estimate and then updates the posterior state and system noise estimations using the latest RSSI observation and current gain value. The general form of the algorithm assumes a linear-Gaussian model denoted as:

\begin{equation}
f(x(k)|x(k-1)) = N(Fx(k-1),Q)
\end{equation}
where the posterior state given the prior is a function of:

\begin{itemize}
\item the system/linear transition matrix $F$ multiplied by the prior state estimation, and
\item the system covariance estimate $Q$.
\end{itemize}
 
Since the proximity measurements are static, velocity is assumed to be zero, and so, the system matrix $F$ is the identity matrix with discretization period $\triangle$ of 100ms/ 0.1s, set by the transmission interval of the BLE beacons, as:

\begin{equation}
F = 
\begin{pmatrix}
    1 & \triangle\\
    0 & 0
\end{pmatrix}
\end{equation}    
   
This means that F can really be represented one-dimensionally and be represented as $F=1$, where each new iteration of the filter takes place every $\triangle = 0.1$ seconds.
  
The important assumption is that the posterior density $f(x(k)|R(k))$ is Gaussian for all $k$. The filter is executed recursively as in~\cite{GuassianKalman} and the process is shown in Fig.~\ref{kalman}:

\begin{figure}[t!]
\centering%
\includegraphics[width=\columnwidth]{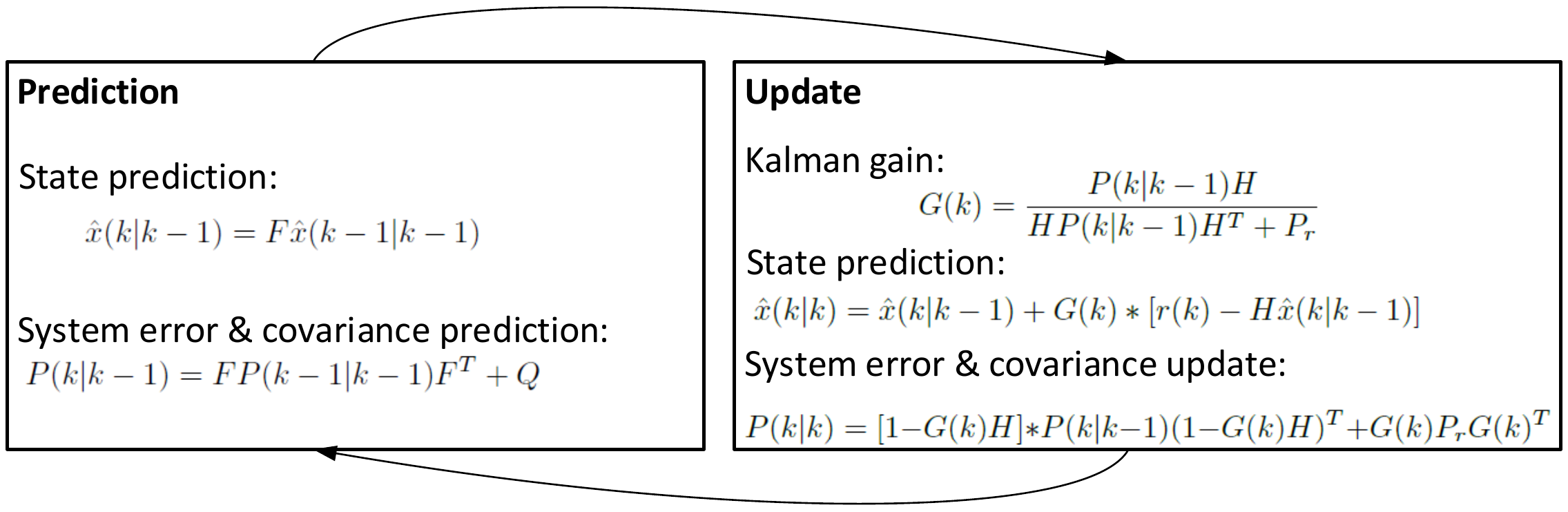}%
\caption{Prediction and update steps in Kalman Filter.}%
\label{kalman}%
\vspace{-3ex}
\end{figure}

%
%
%
%

This system is one dimensional, and thus the covariance matrix $Q$ is just calculated as a variance of a set of previously collected observations $R(k) = \{r(k),r(k-1),...,r(k-N)\}$.  
 
\begin{equation}
Q = \frac{\sum_{i=1}^{N}(r(k) - \mu)^2} {N}
\end{equation}
where $N$ is a constant that defines the size of the set $R$ and $\mu$ is the mean of the observations in the set.

Both the prediction and the update stages are recursive. The prediction stage is needed to make an estimation on the RSSI value based on the previously determined RSSI value and a recursively calculated process noise value, while the update stage computes a gain value that influences the final result. A greater Kalman gain results in a greater influence on the final value. Hence, as the Kalman gain increases, the more influence the predicted value has over the measured value, and so the final result will drift more towards the predicted value.

Our previous work explores the use of a static Kalman filter (KF-ST)~\cite{globalsip},  in which the process noise $Q$ is assumed to be zero when taking stationary proximity measurements. This was assumed due to the static nature of the proximity measurements and the semi-controlled state of each environment. In this work, this assumption is changed in order to account for potential LOS and interference nonlinearities. Thus, a dynamic variation of the KF-ST computes $Q$ at every iteration in order to make up for real-world process noise changes. The initial conditions are defined as $Q(k_0) = 0$. This holds true only for the first RSSI measurement $r(0)$. This initial condition assumes that the model is effective in predicting the state, which we can only test when multiple measurements have been made. However, the system can correct itself quickly due to the frequent transmission intervals.

\begin{figure}[t!]
\centering%
\includegraphics[width=\columnwidth]{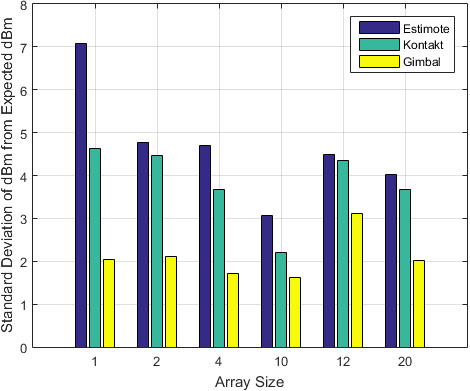}%
\caption{Standard deviation in dBm vs. RSSI set size $N$.}%
\label{arraySize}%
\vspace{-3ex}
\end{figure}

Different set sizes $N$ of RSSI observations $r$ were tested to determine the ideal number set size to utilize in this calculation. The difference was expressed as the standard deviation in dBm. As shown in Fig.~\ref{arraySize}, through experimentation with three popular beacons, Estimote~\cite{estimote}, Kontakt~\cite{kontakt}, and Gimbal~\cite{gimbal}, as the array size increases the precision increases as well, up to a set size $N=10$. The increase in precision limits the effects of outliers and high variation within the set of observed RSSI values. Above 10, any increase of the size leads to a decrease in the precision for all the three beacons. This is likely due to \textit{overfitting}, where the number of values is too high, resulting in poor predictive performance. Hence, size 10 would be ideal for this experimentation. Furthermore, an array size greater than 10 can lead to a waste of resources and limit computational performance since the filter runs on a smartphone, whereas a decrease in the set size below 10 does not provide a sufficient number of observations in order to improve its estimation.

The set of $R$ of RSSI observations is stored in an array list of data type double. Entries are appended to the array starting from index 0. It follows the standard FIFO queue structure. At the start of each iteration, the algorithm checks the size of the array, and once it reaches the desired size 10, it removes the oldest entry (index 0) and adds in the newest measurement. 

The dynamic element to the filter adds a deal of complexity worth mentioning. In comparison to the KF-ST, which only had to complete 5 simple calculations to the order of $O(n)$, the KF-DN must also maintain an array as well as calculate the variance; a function represented as $O(n^2)$ in complexity.


%
%
%
%
%
%
%
%

\subsection{Particle filter}\label{pf}
Particle filter,  is a nonlinear, Bayesian filter well suited for noisy measurements. It can estimate the current state by utilizing all measurement observations $k$ up to and including time $i$. Particle filter is designed and laid out similar to~\cite{particleFilt}. The measurement is represented by a state sequence $\{x_i, i \in N\}$ given by: 

\begin{equation}\label{stateSeq}
x_i = g_i(x_{i-1}|m_{i-1})
\end{equation}
where $g_i$ is a nonlinear state function of the previous state $x_{i-1}$ and  $\{m_{i-1}, i \in N\}$ is an independent, identically distributed process, representing the noise sequence.

Particle filter estimates the current state $x_i$ recursively by utilizing all measurement observations $k_i$ up to and including time $i$. The computational load of each measurement is given as:

\begin{equation}
k_i = h_i(x_i|n_i)
\end{equation}
where $h_i$ is a nonlinear function of the current state $y_i$ and the independent measurement noise sequence $n_i$ of the system.

Let $\{x_{0:i}^j, w_{i}^j\}_{j=1}^{N_s}$  denote a set of  random measurements to characterizes the posterior pdf $p(x_{0:i}|k_{1:i})$, where
$\{x_{0:i}^j, j=0,...,N_s\}$ is the support points set which has associated weights $\{w_i^j, j=1,...,N_s\}$, and $x_{0:i}=\{x_n, n= 0,...,i\}$ is the set of all states up to time $i$. The weights are all normalized at 1 and the posterior density at $i$ is approximately:

\begin{equation}
p(x_{0:i}|k_{1:i}) \approx \sum_{j=1}^{N_{s}}{w_i^j}\delta(x_{0:i} - x_{0:i}^j)
\end{equation}

This is a discrete weighted approximation of the posterior probability. The samples are chosen according to their significancy on the global density. However, we can not concentrate the samples adaptively on the high probability regions in order to obtain variable resolution. Particle filter takes advantage of Monte Carlo numerical integration methods to provide a recursive implementation of the Bayes filter. The weights are chosen using importance sampling~\cite{Bergman}, and the weighted approximation density is:

\begin{equation}
p(x) \approx  \sum_{i=1}^{N_{s}}{w^j}\delta(x - x^j)
\end{equation}
where
\begin{equation}
w(x_j) \propto \frac{\pi(x^j)}{q(x^j)}
\end{equation}
is the normalized weight of the $j$ the particle.

The principle of the importance sampling is used to approximate the posterior distribution as~\cite{particleFilt}:

\begin{equation}
p(x_i|k_{1:i}) \approx \sum_{j=1}^{N_{s}}{w_i^j}\delta(x_i - x_i^j)
\end{equation}

where the weights are:

\begin{equation}
w_i^j \propto  w_{i-1}^j \frac{p(k_i|x_i^j)p(x_i^j|x_{i-1}^j)}{q(x_i^j|x_{i-1}^j), k_i}
\end{equation}

Hence, the algorithm consists of recursive propagation of the weights, since each measurement is observed sequentially.

For all experiments, the number of particles used was 100. This number was chosen based on iterative experimentation, starting with 50 particles and increase by 50 particles up to and include 2000 particles. It was found that 100 particles are optimal in this scenario for the chosen environments, however, it should be reconsidered for any new environment. Also, note that the increased number of particles increases the computational load at the smartphone.

\subsection{Nonparametric Information (NI) filter}\label{nif}
NI is a closed-form version of the standard Bayesian filter. Its design shares similarities with the Kalman filter, however, its purpose is to handle nonlinear and non-Gaussian conditions, which is particularly useful for the non-parametric path loss model. NI in this experiment is modeled following~\cite{wlan}.

First the prediction density  $f(x(k)|R(k-1))$ can be once again described as a Gaussian density, similar to the Kalman filter.

\begin{itemize}
\item \textbf{Prediction:}
\end{itemize}

\begin{equation}
\hat{x}(k|k-1) = F\hat{x}(k-1|k-1) \label{one}
\end{equation}

\begin{equation}
P(k|k-1) = FP(k-1|k-1)F^T + Q \label{two}
\end{equation}

Specifically, NI computes the update stage directly using a joint kernel density estimator (KDE). This combines the contributions of the KDE memoryless estimator with the prediction of the dynamic model. Whereas the Kalman filter predicts the state estimation $f(x(k-1)|R(k-1))$ to be Gaussian, the NI approximates the density with a Gaussian density.

The KDE used in the NI is a Gaussian approximation KDE, where the KDE requires a set of training pairs:

\begin{equation}
 \{(x_i,\bar{r}_i)|i=1,..., N\}
 \end{equation}
 where $x_i$ is the proximity and $\bar{r}_i$ is the average observation RSSI at that proximity, up to $N$ training points. In our experiments, the training points are previously defined by the 14 proximity points $D$. 
 
The general form of the kernel density estimate is given as:

\begin{equation} 
\hat{f}(r|p) = \frac{1}{n\sigma_r}\sum_{t=1}^nK(\frac{r-r(k)}{\Sigma_r}) \label{three}
\end{equation}
where $K()$ is a zero mean, non-negative kernel function, and $\sum_r$ can be thought of as the bin width of a histogram and acts as a smoothing parameter for the KDE. In the case of this application model, the KDE is 1-dimensional. 

For our path loss model, we chose to employ the Gaussian kernel, which is given as:

\begin{equation}
K(r_i) =  \frac{1}{(2\pi)^{\frac{d}{2}}|\Sigma|^{\frac{1}{2}}}exp{(-\frac{(r(k)-\bar{r}_i)^T(r(k)-\bar{r}_i)}{2\Sigma})}
\end{equation}
where the smoothing parameter $\sum=1$, the dimension $d=1$, and $\bar{r}_i$ is the training point (RSSI average) at proximity $x_i$. 

Given the set of RSSI observations $\{r(k)|k=1,..., N\}$ at anchor point $p$, the Gaussian kernel density estimate is now given as:

\begin{equation}
f(x(k)|r(k)) \approx \sum_{i=1}^N\omega_i(r(k))K(x_i)
\end{equation}

with the weights ($\omega_i$) calculated as:

\begin{equation}
\omega_i = \frac{K(r_i)}{\sum_{i=1}^NK(r_i)} 
\end{equation}

The formulation given by Eq. (\ref{one}), (\ref{two}), and (\ref{three}) determines the posterior distribution using a Gaussian kernel; the estimate of a distributed set of RSSI points. This is the defining difference between the NI and the Kalman filter.

The remaining portion of the spatial processing/observation step is to determine the state estimate and observation noise based on the weighting scheme determined from the Gaussian KDE. 

\begin{equation}
\hat{x}_r(k) = \sum_{i=1}^N\omega_i(r(k))x_i 
\end{equation}

\begin{equation}
P_r(k) = \sum_{i=1}^N\omega_i(r(k))(\Sigma_r + (\hat{x}_r(k)-x_i)(\hat{x}_r(k)-x_i)^T)
\end{equation}

Once $\hat{x}_r(k)$ and $P_r(k)$ are computed, the posterior state density and process noise density can be updated.

\begin{itemize}
\item \textbf{Update (Correction):}
\end{itemize}

\begin{equation}
\hat{x}_r = P(k|k)(P(k|k-1)^{-1}\hat{x}(k|k-1)+P_r^{-1}\hat{x}_r(k))
\end{equation}

\begin{equation}
P(k|k)^{-1} = P(k|k-1)^{-1}+P_r^{-1}(k)
\end{equation}

Note that the initial conditions of the posterior state and process noise densities are set to zero.

\section{Experimental Setup}\label{system}
In this section, the system components and the experimental set up are described. The equipment and the experimental environment are presented, followed by the wireless communication protocol and the experimental procedure.

\begin{figure}[t]
\centering%
\includegraphics[width=\columnwidth]{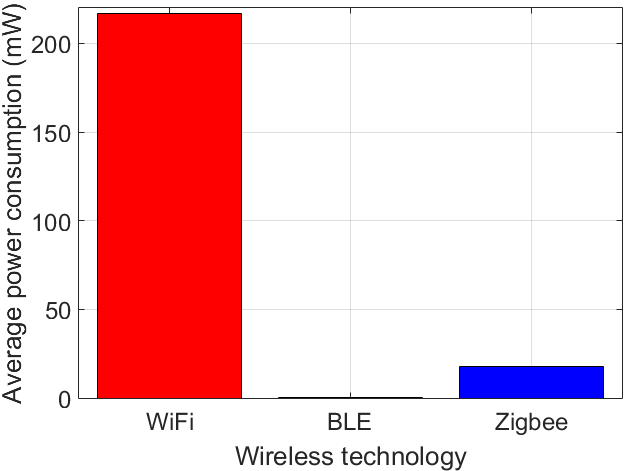}%
\caption{Average power consumption of wireless technologies.}%
\label{comp}%

\end{figure}

Since energy requirement is important in this study, initially, we examine the average power consumption of two other popular wireless technologies; WiFi (2.4 GHz) and ZigBee. The results are shown in Fig.~\ref{comp}. It is clear that BLE has the lowest power consumption for PBS.

\subsection{Equipment}
There is a huge variety of BLE beacon devices on the market. Each has its own unique features, such as additional sensors, battery life, reconfigurability, and dimensions, though all fundamentally work the same. Depending on the application, it may be important to choose wisely. For the purposes of this experiment and to examine beacons with different hardware, beacons from three manufacturers were used to establish more conclusive results: Estimote~\cite{estimote}, Kontakt~\cite{kontakt}, and Gimbal~\cite{gimbal}. The specifications of the beacons are shown in Table~\ref{beaconspecs}, along with the estimated battery lifetime of each beacon, as it was calculated after experimentation. The receiving and calculating device was a Google Nexus~5, running Android version~6.0.1 and Bluetooth~4.0. An Android application that uses the \texttt{AltBeacon} Android library is utilized to interact with the BLE beacons. Finally, a digital laser distance finder was used to provide accurate distance placements during the experiments.

It is important to mention that the three beacons were selected in order to examine the performance of the three different chipsets. As beacons becoming popular in many smart city applications, different BLE chipsets are developed, in order to handle the necessary data transmission and the more sensors which are added to the beacons, such as light, temperature, etc. Each chipset has different current consumption, however, for this experimentation, only single data transmission was used for each beacon. This is the main functionality needed for PBS applications to find the distance between the beacon and the user. However, for different applications, the power consumption of the beacon chipset should be carefully examined.

\begin{table}
\centering%
\small%
\renewcommand{\arraystretch}{1.75}%
\setlength{\tabcolsep}{0.24\tabcolsep}%
\newlength{\cellwdt}%
\setlength{\cellwdt}{0.25\columnwidth}%
\addtolength{\cellwdt}{-2\tabcolsep}%
\addtolength{\cellwdt}{-0.6pt}%
\caption{Specifications of the beacons.}%

\begin{tabular}{|l|c|c|c|}\hline

\makebox[\cellwdt][l]{\multirow{2}{*}\textbf{\textbf{Vendor}}} & \includegraphics[width=0.9cm]{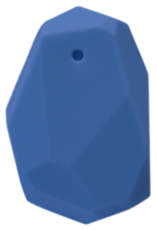}
 & \includegraphics[width=1cm]{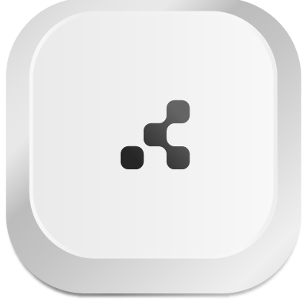}
 & \includegraphics[width=0.9cm]{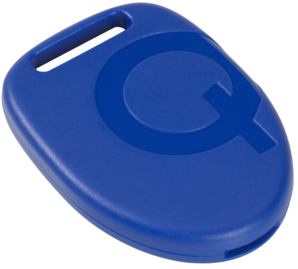}
\\ 
 & \makebox[\cellwdt][c]{\textbf{Estimote}~\cite{estimote}} & \makebox[\cellwdt][c]{\textbf{Kontakt}~\cite{kontakt}} & \makebox[\cellwdt][c]{\textbf{Gimbal}~\cite{gimbal}}\\ \hline
\textbf{Chipset} &  nRF2832& nRF51822 &  Gimbal \\ \hline
\textbf{Power Supply} &4 x CR2477  &   2 x CR2477 & 1 x  CR2032\\ \hline
\textbf{Est. Lifetime} & 21.4 months & 16.4 months & 0.9 months \\ \hline

\end{tabular}

\label{beaconspecs}%

\end{table}



\subsection{Experimental environment}\label{topology}

\begin{figure}[t]
\centering
\captionsetup[subfloat]{farskip=0pt}
\subfloat[Environment 1: Large room.]{\includegraphics[width=0.44\columnwidth]{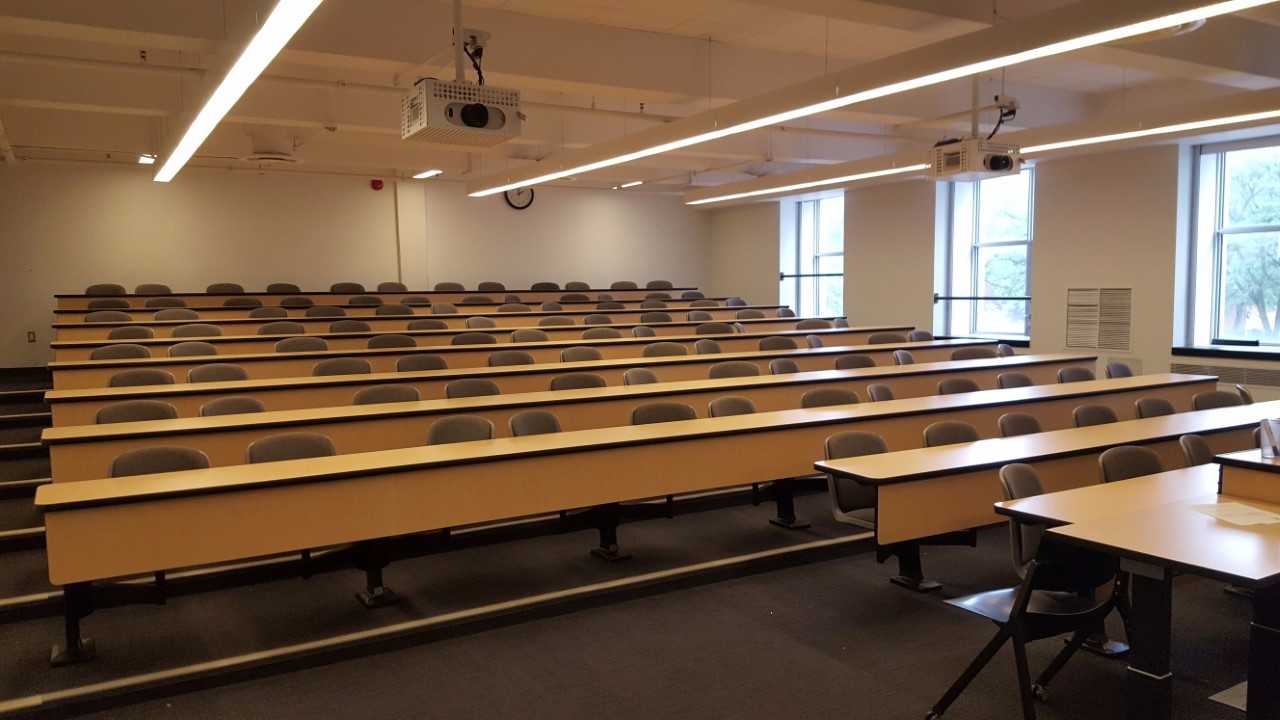}%
\label{largeRoom}
}\quad \quad
\subfloat[Environment 2: Small room.]
{\includegraphics[width=0.44\columnwidth]{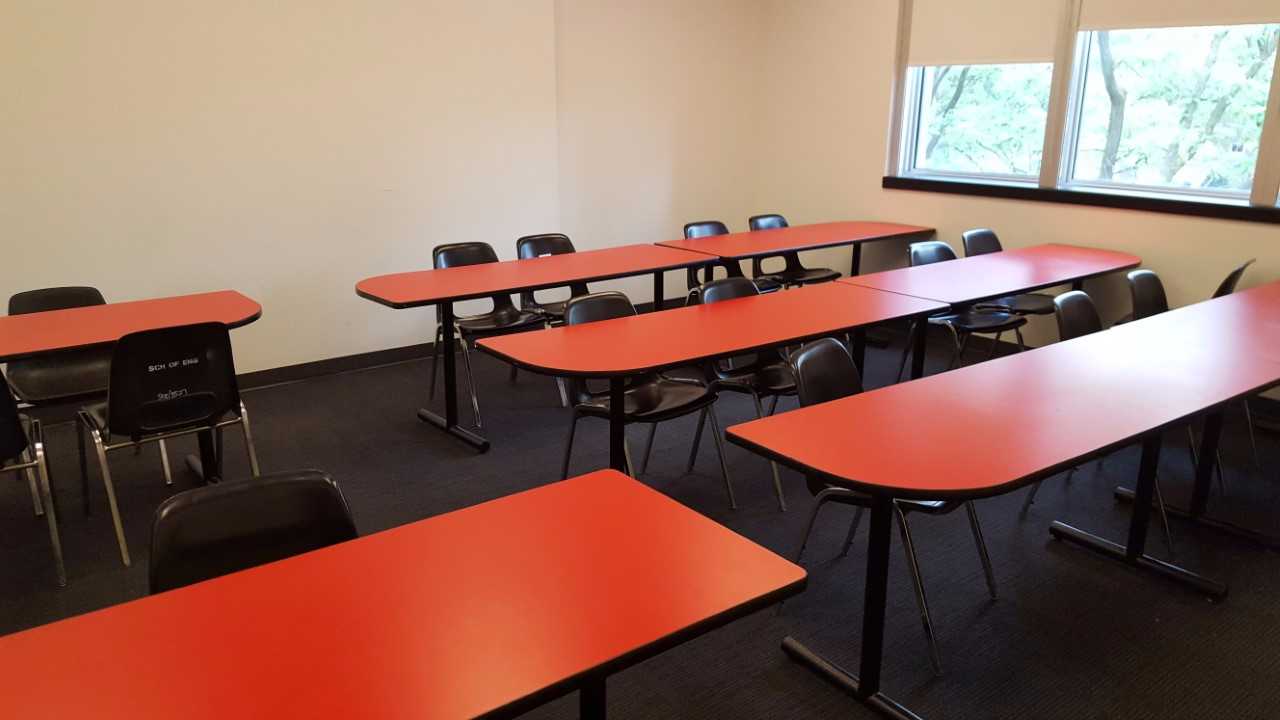}%
\label{smallRoom}
}
\caption{Experimental environments (a) a large room (11m~$\times$~9m) and (b) a small room (6m~$\times$~4m).}\label{expenv}
\vspace{-3ex}
\end{figure}

To capture the behavior and performance of all the filters, the experiments were conducted in two rooms of various sizes. The first, a large lecture hall with dimensions 11m $\times$ 9m, shown in Fig.~\ref{largeRoom}. The beacons were placed in 1~m height from the floor, while there were no people in the room during experimentation, and only wooden desks and plastic chairs were in the room. The second, a smaller meeting room with dimensions 6m $\times$ 4m, shown in  Fig.~\ref{smallRoom}, the beacons again were placed in 1~m height. No people were in the room during experimentation and there were only plastic desks and chairs in the room. The temperature during experimentation in both rooms was $25^\circ$C, while metal objects, plants and any other objects that could affect the performance of the beacons were removed from each room, for the experiment. Due to the sensitivity of the RSSI signals on the environmental parameters, changes in the above parameters might affect the final results.

 Both rooms were laid out with a set of tables and chairs. The rooms are chosen as they present common size and layout characteristics to many indoor environments. Furthermore, the furniture layout defines a common set of physical objects found in many indoor environments. The two rooms were selected to capture sufficient properties and qualities common to many indoor environments.

\newcommand{\gscale}{0.46}
\begin{figure*}[t!] 
\centering
\captionsetup[subfloat]{farskip=0pt}%
\subfloat[Environment 1 - Estimote.]{\includegraphics[scale=\gscale]{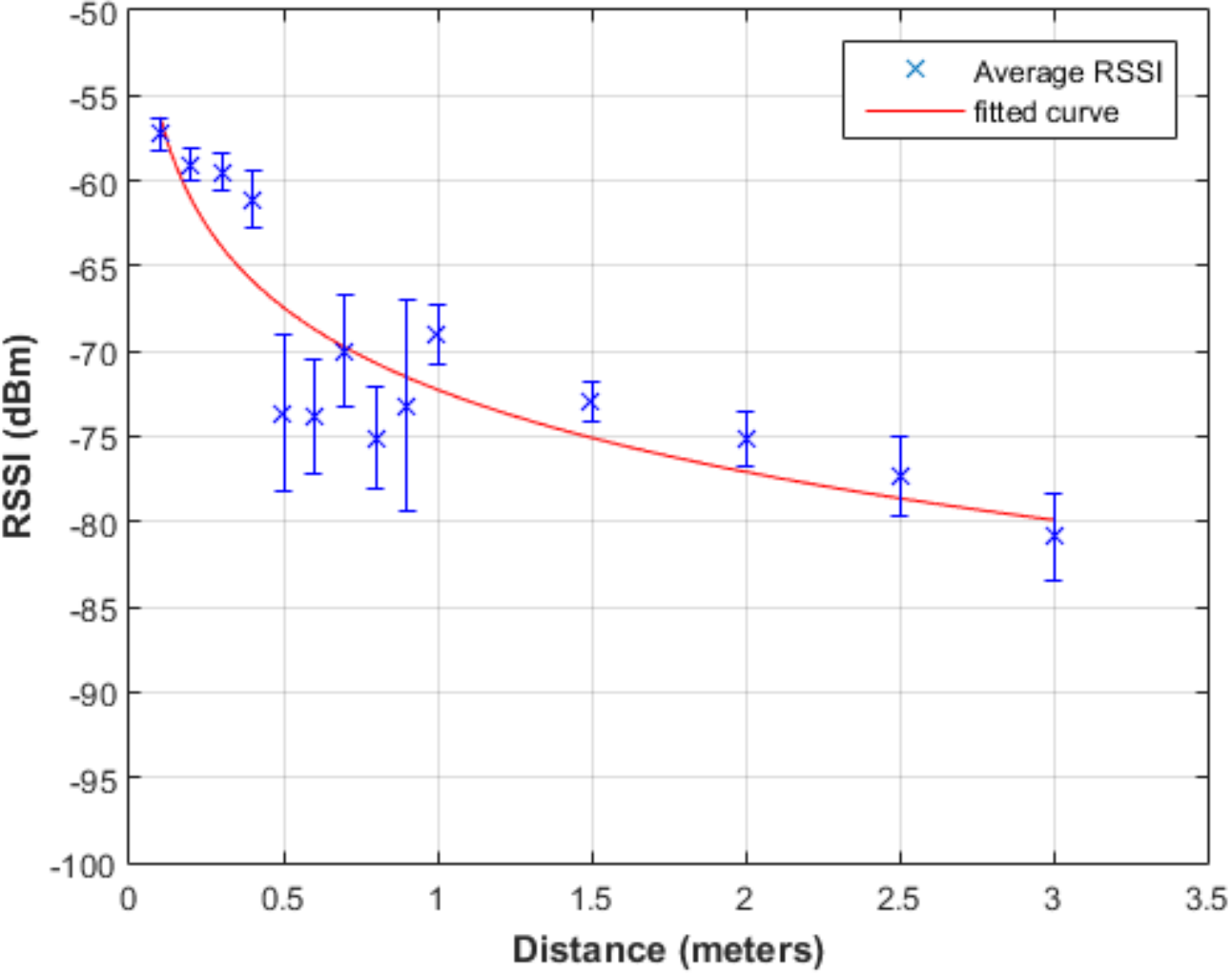}}
\subfloat[Environment 1 - Kontakt.]{\includegraphics[scale=\gscale]{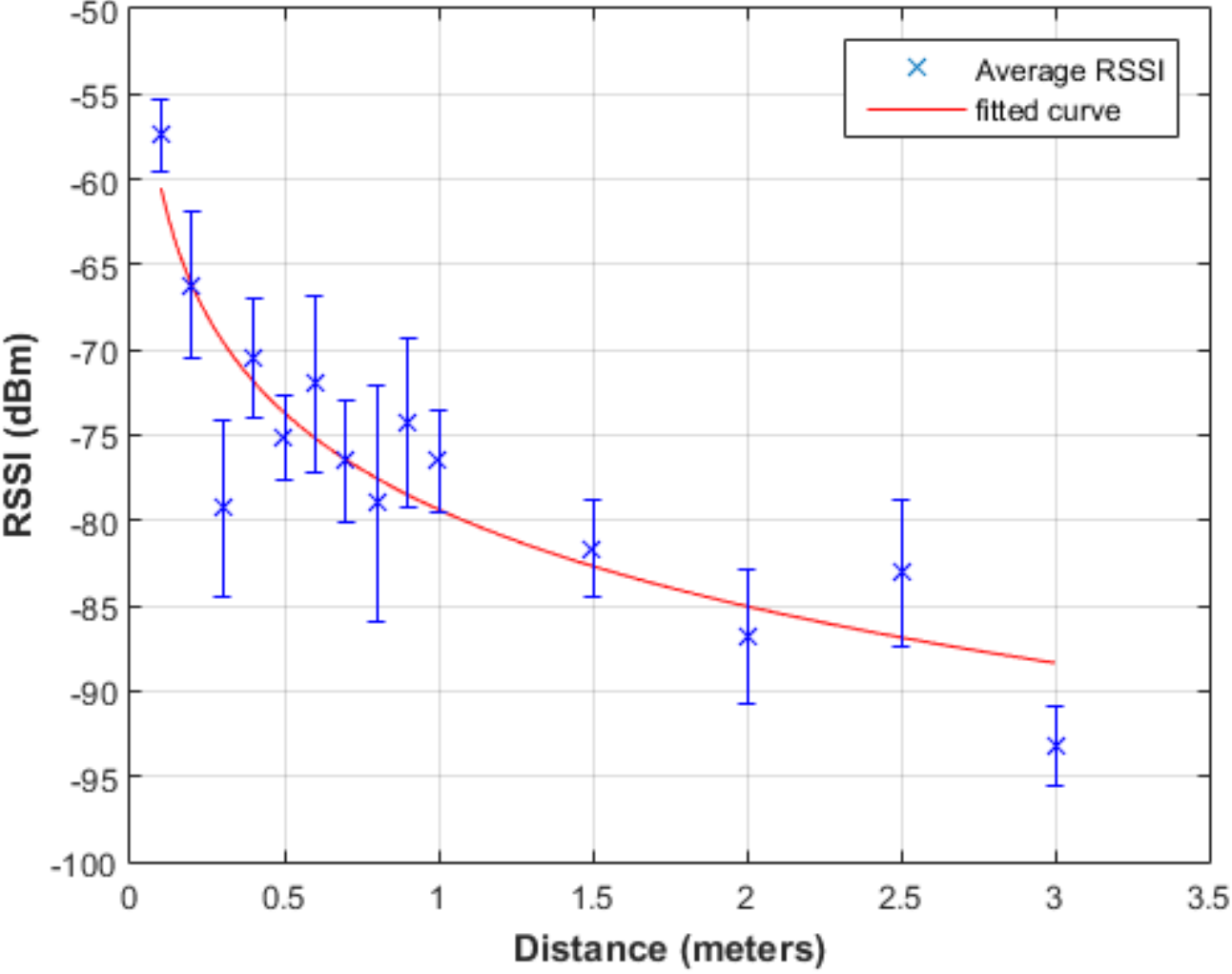}}
\subfloat[Environment 1 - Gimbal.]{\includegraphics[scale=\gscale]{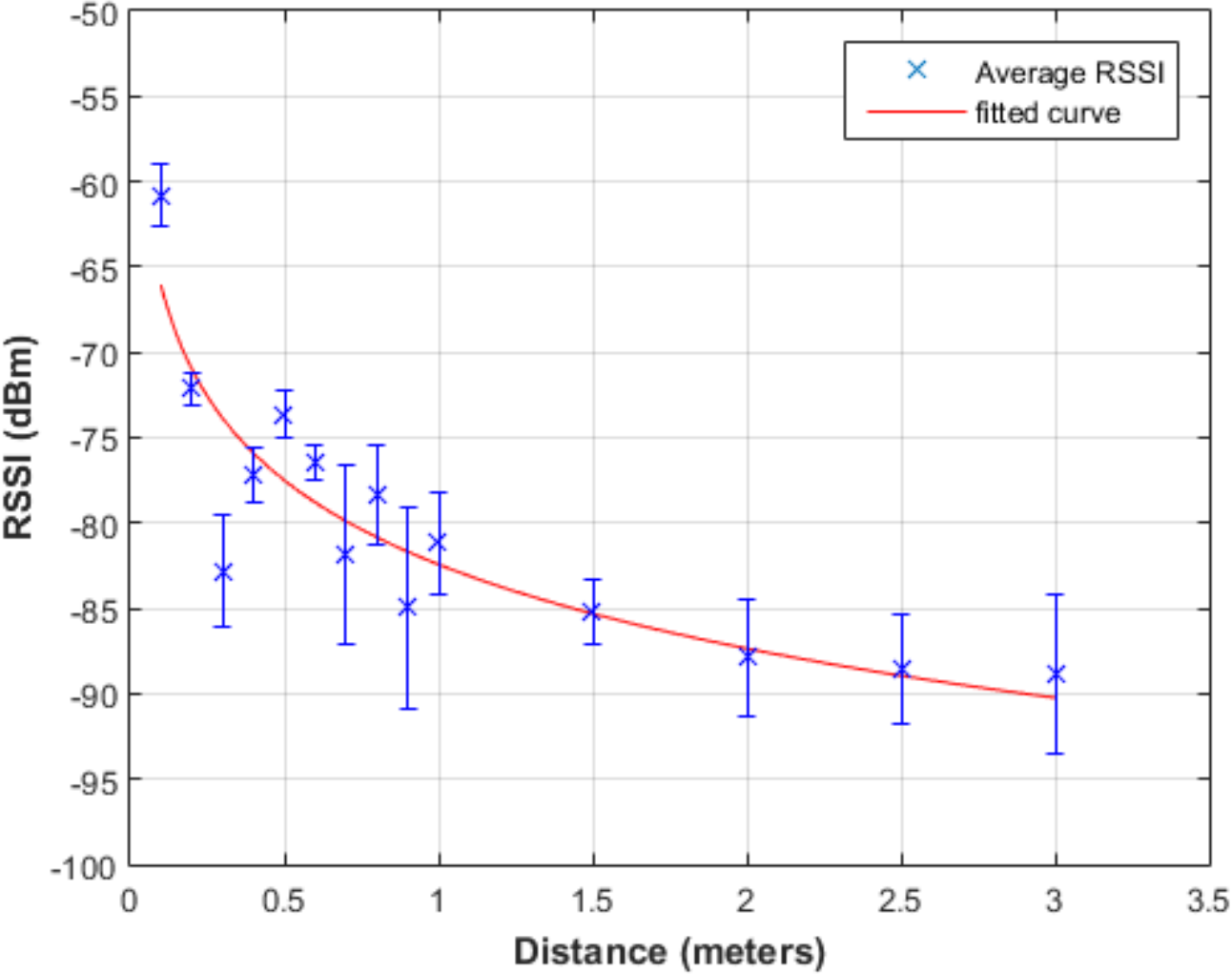}}
\caption{Environment 1 (11m $\times$ 9m): Path-loss model.}
\label{r1PL}%
\end{figure*}

\begin{figure*}[t!]
\captionsetup[subfloat]{farskip=0pt}%

\subfloat[Environment 2 - Estimote.]{\includegraphics[scale=\gscale]{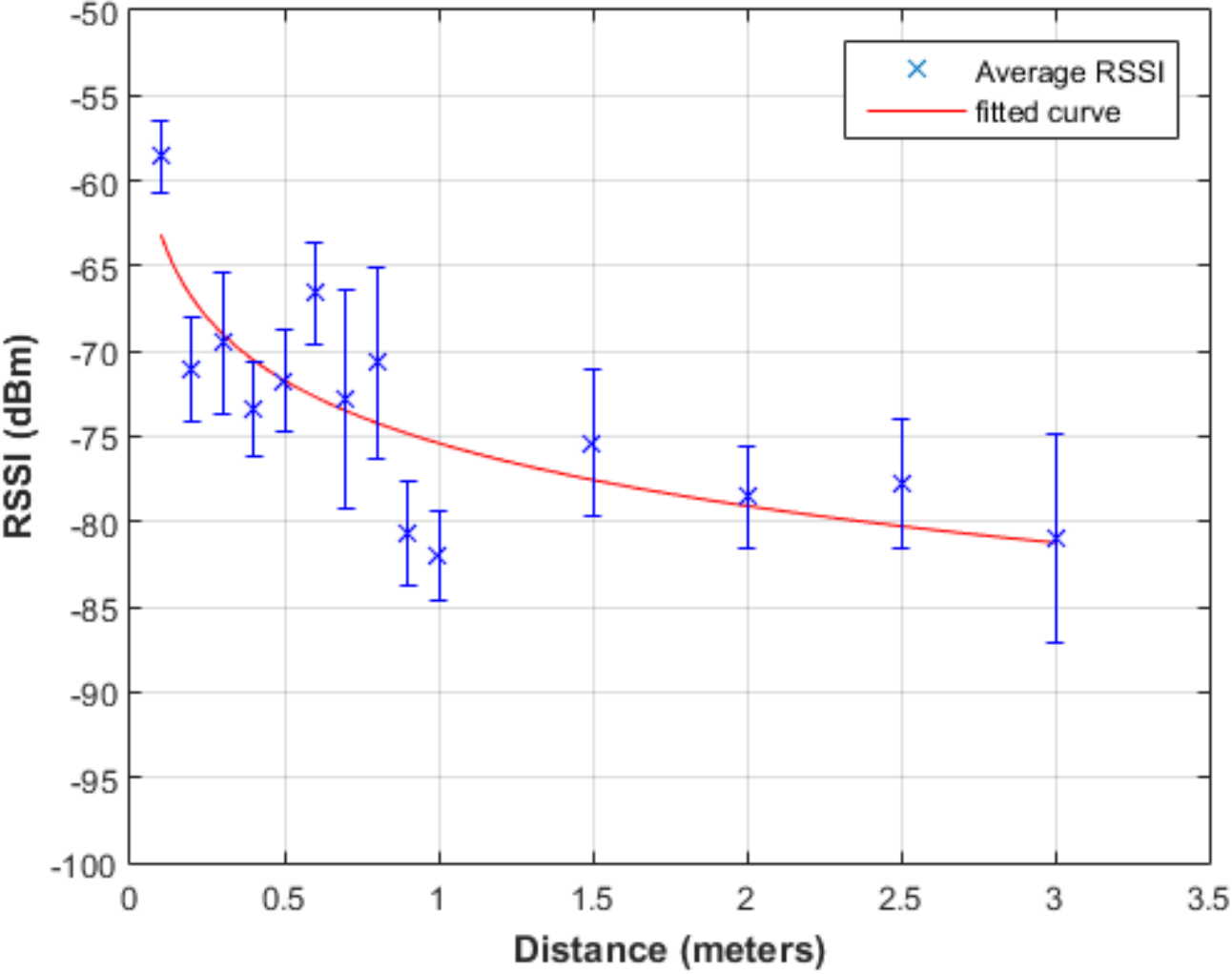}}
\subfloat[Environment 2 - Kontakt.]{\includegraphics[scale=\gscale]{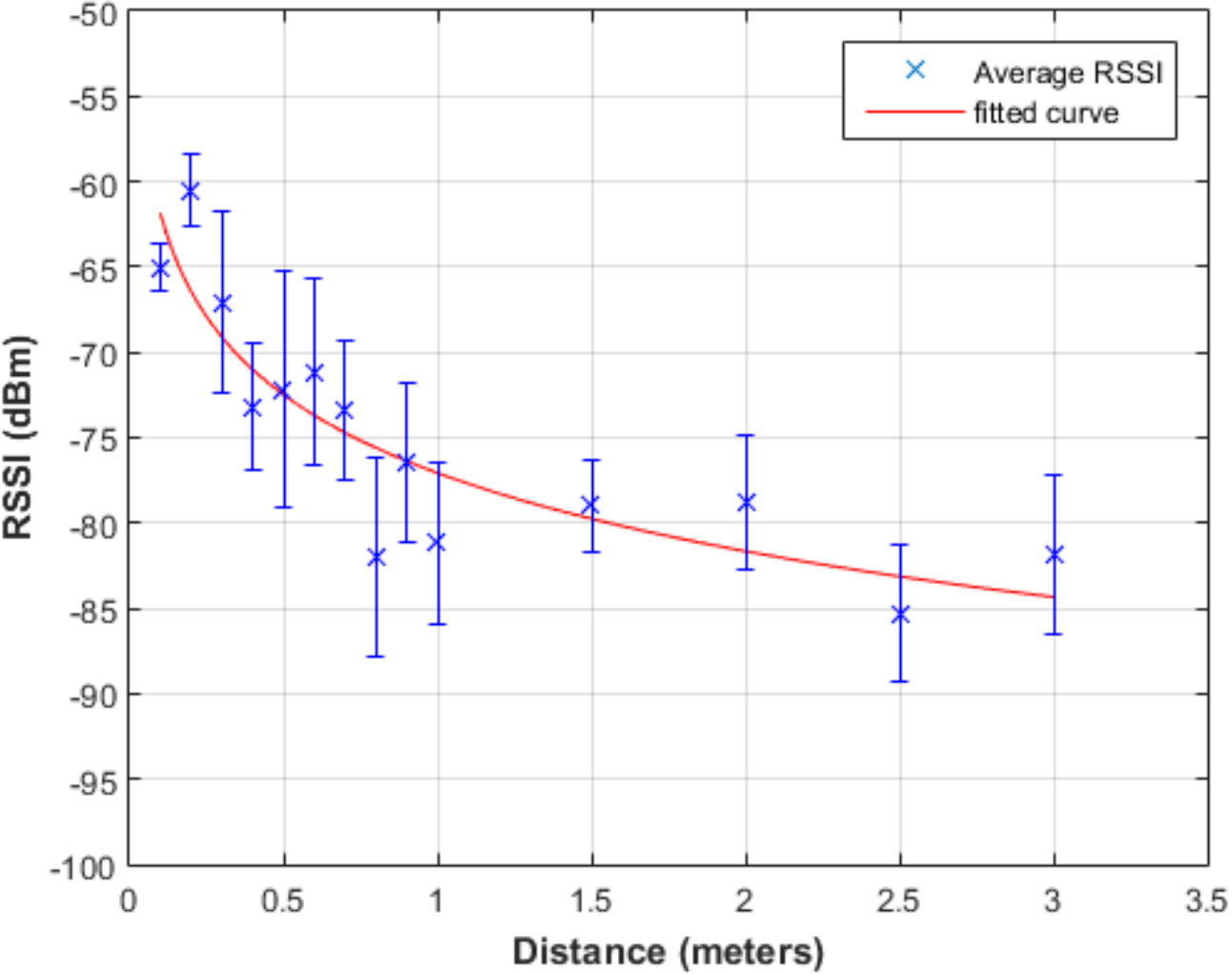}}
\subfloat[Environment 2 - Gimbal.]{\includegraphics[scale=\gscale]{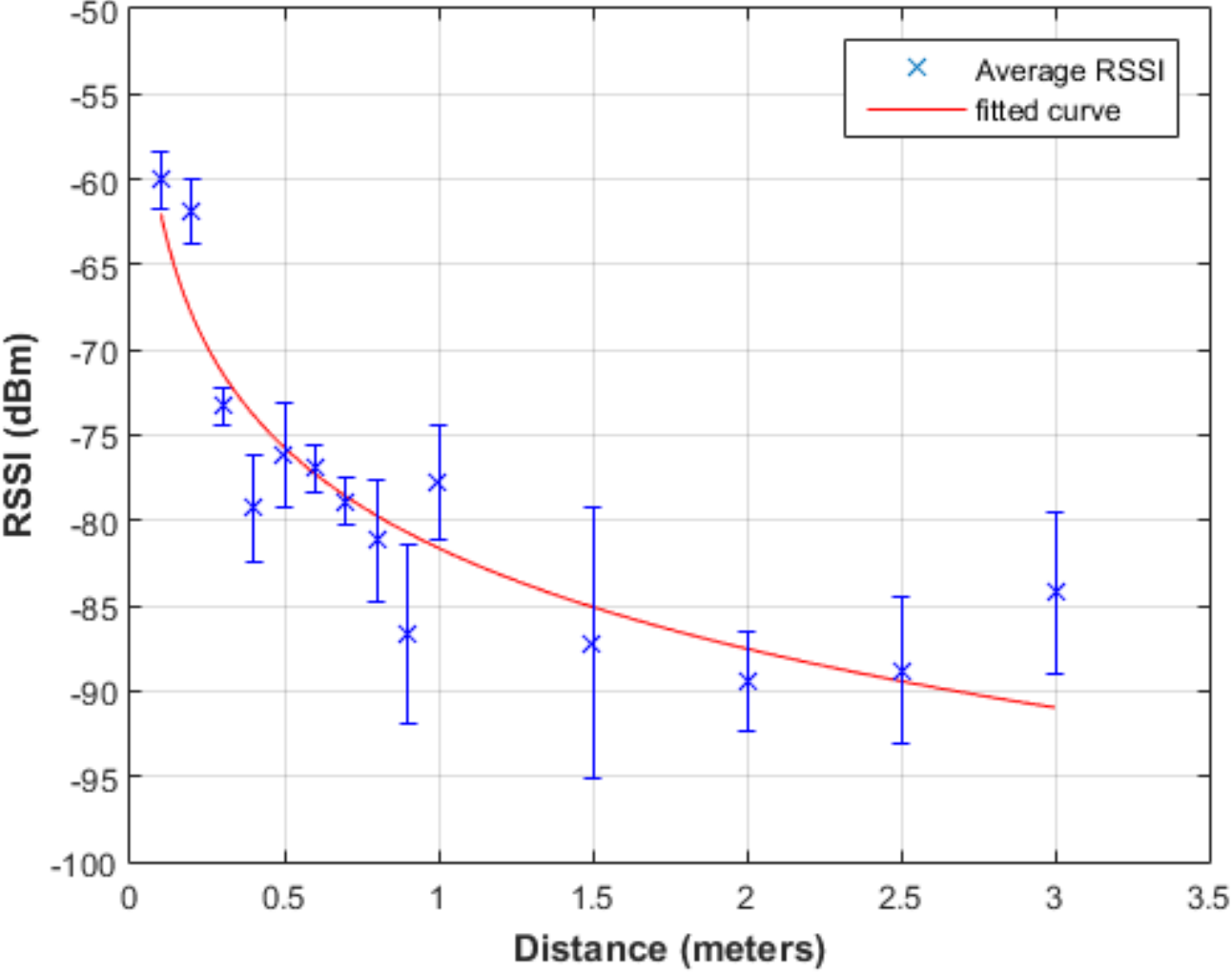}}
\caption{Environment 2 (6m $\times$ 4m): Path-loss model.}
\label{r2PL}%
\vspace{-3ex}
\end{figure*}

The variation in the two room sizes provides contrasting environmental conditions to better compare and validate the results and determine trends. Control over local environmental settings during the experiments is difficult, as small changes in the environment over time may have a large impact on beacon performance. To mitigate noise, particularly due to physical obstructions, no objects were added or removed from each room (i.e. chairs, tables, etc.), nor was anybody granted access to the rooms for the duration of all experiments.

\subsection{Communication protocol}\label{comm}
Most beacons implement the standard communication protocols: iBeacon and Eddystone, set by Apple and Google respectively.  In this experiment, we used Apple's iBeacon protocol. The iBeacon protocol was introduced in 2013 and defines a packet with three distinct configurable fields: a 16-byte universal unique identifier (UUID), a 2-byte major value, and 2-byte minor value~\cite{ibeaconpacket}. These fields are used to subregion, and subsection the devices in a unique way, as per the user-developed application. All beacons that implement this protocol must have these three fields, but each manufacturer may have additional fields in their packet structure. For instance, the Kontakt beacon has a total of 30 bytes make up the full packet. Other fields transmit useful data such as manufacturer information (bytes 5-9)~\cite{kontaktpacket}. 

For this experiment, all beacons used the packet structures defined by their respective manufacturer but were all configured to have a transmission power of -12~dBm, transmitting at an interval of 100~ms. Note that the receiving device (Google Nexus 5) is using a sampling time of 11.25~ms as defined by the Google Bluetooth stack for Android devices~\cite{googleBLE}. This is far more frequent than the 100~ms transmission interval, making the system robust against packet loss.

%

\subsection{Experimental Procedure}\label{procedure}
At the beginning of each experiment, the path loss for each of the experimental environment was calculated. For every beacon, approximately 1000 data were collected. The data was used in order to calibrate the $C_0$ and $n$ values, according to the unique hardware characteristics of each beacon.

Then, a simple procedure was developed to get a suitable set of results for each beacon. The procedure runs for each of the five types of filters; SMA, KF-ST, KF-DN, NI, and PF. 

\begin{itemize}
\item The first part of the experiment obtains 1000 raw RSSI data points at each distance. 
\item The second part of the experiment applies the filtering technique.  
\end{itemize}

Within each iteration, 14 measurement points are taken at 2 minutes each. With a transmission interval of 100~ms, this leads to a total of approx. 1200  data points per position and approx. 16800 data points per beacon, per environment. 

The beacon remains stationary on a table, while the smartphone moves along the table at the following set of distances:
  
$$D = \{0.1, 0.2, 0.3, 0.4, 0.5, 0.6, 0.7,$$$$ 0.8, 0.9, 1.0, 1.5, 2.0, 2.5,3.0\}(meters)$$

At each distance, the RSSI and corresponding distance are recorded.

\section{Experimental Results and Discussion}\label{exp}
In this section, the experimental results are described followed by a brief discussion.
\renewcommand{\arraystretch}{1}%
\newlength{\cellwidtha}%
\newlength{\cellwidthb}%

\begin{table}[t!]
\centering
\renewcommand{\arraystretch}{1}%
\caption{Path-loss values.}%
\begin{tabular}{|l|c|c|c|c|c|c|}
\hline
\multirow{2}{*}{} & \multicolumn{2}{c|}{Estimote} & \multicolumn{2}{c|}{Kontakt} & \multicolumn{2}{c|}{Gimbal} \\ \cline{2-7} 
                  &   $C_0$            &   n          &    $C_0$             &   n          &    $C_0$            &   n           \\ \hline
Environment 1         & -72.25         & 1.601        & -79.35        & 1.885       & -82.42        & 1.960  \\ \hline
Environment 2          & -75.39         & 1.224       & -77.07        & 1.523        & -81.61        & 1.637       \\ \hline
\end{tabular}
\label{roomPL}
\end{table}

\begin{table*}[t!]
\caption{Environment 1: Absolute Error for all distances (\textnormal{m}), for Estimote (\textit{E}), Kontakt (\textit{K}) and Gimbal (\textit{G}) beacon.}
\resizebox{\textwidth}{!}{%
\begin{tabular}{|l|c|c|c|c|c|c|c|c|c|c|c|c|c|c|c|c|c|c|}
\hline
\textbf{} & \multicolumn{3}{c|}{\textbf{0.5m}} & \multicolumn{3}{c|}{\textbf{1.0m}} & \multicolumn{3}{c|}{\textbf{1.5m}} & \multicolumn{3}{c|}{\textbf{2.0m}} & \multicolumn{3}{c|}{\textbf{2.5m}} & \multicolumn{3}{c|}{\textbf{3.0m}} \\ \hline
\textbf{SMA} & 0.38 & 0.01 & 0.19 & 0.40 & 0.43 & 0.24 & 0.32 & 0.35 & 0.01 & 0.24 & 0.43 & 1.59 & 0.15 & 1.51 & 0.64 & 1.69 & 2.08 & 1.40 \\ \hline
\textbf{KF-ST} & 0.69 & 0.09 & 0.22 & 0.39 & 0.31 & 0.18 & 0.41 & 0.20 & 0.06 & 0.51 & 0.42 & 0.07 & 0.47 & 0.97 & 0.21 & 0.39 & 2.20 & 0.63 \\ \hline
\textbf{KF-DN} & 0.72 & 0.09 & 0.19 & 0.38 & 0.33 & 0.44 & 0.41 & 0.54 & 0.14 & 0.49 & 0.47 & 0.96 & 0.45 & 1.49 & 1.23 & 0.38 & 1.33 & 0.29 \\ \hline
\textbf{NI} & 0.57 & 0.09 & 0.23 & 0.34 & 0.30 & 0.17 & 0.34 & 0.13 & 0.03 & 0.59 & 0.44 & 0.09 & 0.38 & 0.92 & 0.37 & 0.16 & 2.22 & 1.06 \\ \hline
\textbf{PF} & 0.20 & 0.35 & 0.20 & 0.37 & 0.27 & 0.91 & 0.78 & 0.12 & 0.03 & 0.52 & 0.48 & 0.01 & 0.42 & 1.12 & 0.02 & 1.06 & 0.25 & 0.48 \\ \hline
\textbf{Beacon} & \textit{E} & \textit{K} & \textit{G} & \textit{E} & \textit{K} & \textit{G} & \textit{E} & \textit{K} & \textit{G} & \textit{E} & \textit{K} & \textit{G} & \textit{E} & \textit{K} & \textit{G} & \textit{E} & \textit{K} & \textit{G} \\ \hline
\end{tabular}%
}
\end{table*}

\begin{table}[t]
\centering%
\scriptsize%
\renewcommand{\arraystretch}{1}%
\setlength{\tabcolsep}{0.2\tabcolsep}%
\settowidth{\cellwidtha}{Display resolution (pixels)}%
\setlength{\cellwidthb}{\columnwidth}%
\addtolength{\cellwidthb}{-\cellwidtha}%
\addtolength{\cellwidthb}{-2\tabcolsep}%
\setlength{\cellwidthb}{0.23\cellwidthb}%
\addtolength{\cellwidthb}{-2\tabcolsep}%
\caption{Environment 1: MAE (\textnormal{m}).}%
\begin{tabular}{|l|c|c|c|c|c|c|} \hline 
\makebox[\cellwidthb][l] & \makebox[\cellwidthb][c]{SMA} & \makebox[\cellwidthb][c]{KF-ST} & \makebox[\cellwidthb][c]{KF-DN}& \makebox[\cellwidthb][c]{NI}& \makebox[\cellwidthb][c]{PF}\\ \hline \hline
Estimote & 0.400 & 0.343 & 0.334 & \textbf{0.315} & 0.331\\ \hline
Kontakt  & 0.518 & 0.405 & 0.403 & 0.425 & \textbf{0.260}\\ \hline
Gimbal   & 0.412 & 0.242 & 0.367 & 0.282 & \textbf{0.218}\\ \hline \hline
\textit{Average}  & \textit{0.443} & \textit{0.33} & \textit{0.368} & \textit{0.341} & \textit{\textbf{0.27}}\\ \hline
\end{tabular}%
\label{meanTable_r1}%
\end{table}

\begin{table}[t!]
\centering%
\scriptsize%
\renewcommand{\arraystretch}{1}%
\setlength{\tabcolsep}{0.2\tabcolsep}%
\settowidth{\cellwidtha}{Display resolution (pixels)}%
\setlength{\cellwidthb}{\columnwidth}%
\addtolength{\cellwidthb}{-\cellwidtha}%
\addtolength{\cellwidthb}{-2\tabcolsep}%
\setlength{\cellwidthb}{0.23\cellwidthb}%
\addtolength{\cellwidthb}{-2\tabcolsep}%
\caption{Environment  1: RMSE (\textnormal{m}).}%
\begin{tabular}{|l|c|c|c|c|c|c|} \hline
\makebox[\cellwidthb][l] & \makebox[\cellwidthb][c]{SMA} & \makebox[\cellwidthb][c]{KF-ST} & \makebox[\cellwidthb][c]{KF-DN}& \makebox[\cellwidthb][c]{NI} & \makebox[\cellwidthb][c]{PF}\\ \hline \hline
Estimote & 0.590 & 0.415 & 0.408 & \textbf{0.376} & 0.423\\ \hline
Kontakt  & 0.782 & 0.695 & 0.599 & 0.705 & \textbf{0.310}\\ \hline
Gimbal   & 0.653 & 0.323 & 0.528 & 0.398 & \textbf{0.291}\\ \hline \hline
\textit{Average}   & \textit{0.675} & \textit{0.478} & \textit{0.512} & \textit{0.493} & \textit{\textbf{0.341}}\\ \hline 

\end{tabular}%
\label{rmseTable_r1}%
\end{table}

\subsection{Experimental Results}\label{results}
\subsubsection{Path loss model} 
The first step is to obtain RSSI values at multiple defined points. RSSI values at 14 specified positions are gathered. RSSI values are gathered for 2~minutes, resulting in approximately 1000 data-points for each distance. In this case, the values are averaged, producing a set of 14 RSSI values, representative of the approximate path loss model for each room. Next, each set of values is subject to the curve fitting. The $x$- axis data is chosen to be the expected distance/ proximity, while the $y$- axis data is chosen to be the obtained RSSI values for the given environment. The standard path loss formula, shown in Eq.(\ref{equation}), is used to solve for the optimal $C_0$ and $n$. The path loss model for each room can be seen in Fig.~\ref{r1PL}  and Fig.~\ref{r2PL}, for Environment~1 and Environment~2, respectively, while Table ~\ref{roomPL} shows the average $C_0$ and $n$ values for each environment. These values will be used later on for the proximity estimation experiments.

The factory value of $C_0$ (measured power at 1 m distance) for the beacons at an indoor environment is $-70$. However, after the path loss experiment, it is clear that calibration in the environment that the beacons will be used is necessary. Also, it is clear that due to the different hardware characteristics of each beacon, although their $C_0$ and $n$ values are similar, they are not the same. Hence, path loss calculation and further calibration are necessary, before any further experimentation. Regarding the environments, in the small room beacons experience greater interference, probably due to the increased reflections and diffractions occur.

\subsubsection{Environment 1 - Large room}
The absolute error for every distance in Environment 1 is shown in Table~\ref{env1all},  and the mean absolute error and root mean square error can be found in Table~\ref{meanTable_r1} and Table~\ref{rmseTable_r1}, respectively. 

When the distance between the beacon and the receiver is small, within 1.5 m, all the approaches have similar performance. As the distance increases, NI and particle filtering have better performance in comparison with the other techniques, and for all the beacons.
    
Referring to the accuracy measure defined by the MAE shown in Table~\ref{meanTable_r1}, the Bayesian filters always outperform the SMA. Specifically, in comparison with SMA, the KF-ST filter improves the accuracy by 14.25\%, when Estimote is used, by 21.81\% when Kontakt is used and 41.26\%, when Gimbal is used, (25.56\% on average for all the three). The KF-DN filter improves the accuracy by 16.5\%, 22.2\%, and 10\% for Estimote, Kontakt and Gimbal respectively (16.99\% on average), and in comparison with SMA. The NI filter improved the accuracy by 21.25\%, 17.95\%, and 31.55\% for Estimote, Kontakt and Gimbal respectively (23.25\% on average), and in comparison with SMA. The highest improvement is with particle filter which improves the accuracy by 17.25\%, 49.8\%, and 47.08\% for Estimote, Kontakt and Gimbal respectively (39.17\% on average), and in comparison with SMA. Overall, when the average of all the beacons and all the distances is calculated, particle filtering has the best proximity estimation with MAE of 0.27 m, when the distance between the beacon and the receiver is within 3 m.

With regards to the accuracy in terms of RMSE shown in Table~\ref{rmseTable_r1}, again the Bayesian filters outperform the SMA. On average, KF-ST improves the accuracy by 29.18\%,  KF-DN by 24.14\%, NI by 26.96\% and particle by 49.48\%, in comparison with SMA.

\subsubsection{Environment 2 - Small room}
The mean absolute error and root mean square error can be found in Table~\ref{meanTable_r2} and  Table~\ref{rmseTable_r2}, respectively.

It can be seen again that the Kalman, NI, and particle filter consistently outperform the SMA. As the distance between the beacon and the receiver increase, again NI and particle filter tend to have better performance, in terms of accuracy than the other approaches. However, the proximity accuracy in this environment is lower than in the previous environment. This might be due to the small size of the second environment which creates more interference and reflections that can affect the performance of beacons. Hence, filtering is necessary for such an environment to improve proximity accuracy.

Since the environment is smaller and creates higher interference for the beacons, it is expected that filtering will be important. This is clear in Table~\ref{meanTable_r2}, where the MAE is shown. The improvement for KF-ST is 27.86\%, for KF-DN is 27.64\%, for  NI is 30.11\% and for particle 30.61\%, in comparison with SMA. 

Similarly, in Table~\ref{rmseTable_r2} the RMSE is shown. The improvement for KF-ST is 27.62\%, for KF-DN is 25.74\%, for  NI is 31.67\% and for particle 29.46\%, in comparison with SMA. It is interesting to notice that, in the second environment, NI and particle filters have the best performance, with NI filter having the lowest RMSE  of 0.63 m, when the distance between the beacon and the receiver is within 3 m.

\begin{table*}[t!]
\caption{Environment 2: Absolute Error for all distances (\textnormal{m}), for Estimote (\textit{E}), Kontakt (\textit{K}) and Gimbal (\textit{G}) beacon.}
\resizebox{\textwidth}{!}{%
\begin{tabular}{|l|c|c|c|c|c|c|c|c|c|c|c|c|c|c|c|c|c|c|}
\hline
\textbf{} & \multicolumn{3}{c|}{\textbf{0.5m}} & \multicolumn{3}{c|}{\textbf{1.0m}} & \multicolumn{3}{c|}{\textbf{1.5m}} & \multicolumn{3}{c|}{\textbf{2.0m}} & \multicolumn{3}{c|}{\textbf{2.5m}} & \multicolumn{3}{c|}{\textbf{3.0m}} \\ \hline
\textbf{SMA} & 0.12 & 0.68 & 0.17 & 4.10 & 0.58 & 0.49 & 0.36 & 0.44 & 0.79 & 0.40 & 0.76 & 0.32 & 1.71 & 1.02 & 0.31 & 2.45 & 1.58 & 1.90 \\ \hline
\textbf{KF-ST} & 0.01 & 0.03 & 0.01 & 2.38 & 0.83 & 0.37 & 0.52 & 0.17 & 0.39 & 0.23 & 0.72 & 0.42 & 0.98 & 0.93 & 0.23 & 0.23 & 0.95 & 1.67 \\ \hline
\textbf{KF-DN} & 0.01 & 0.21 & 0.02 & 2.39 & 0.37 & 0.36 & 0.52 & 0.45 & 0.41 & 0.23 & 0.17 & 0.46 & 0.97 & 0.65 & 0.29 & 0.22 & 1.53 & 1.69 \\ \hline
\textbf{NIF} & 0.03 & 0.02 & 0.00 & 1.94 & 0.94 & 0.38 & 0.51 & 0.10 & 0.18 & 0.17 & 0.78 & 0.22 & 0.93 & 0.34 & 0.47 & 0.96 & 0.96 & 1.69 \\ \hline
\textbf{PF} & 0.01 & 0.10 & 0.08 & 2.46 & 1.85 & 0.35 & 0.53 & 2.36 & 0.18 & 1.47 & 1.86 & 0.12 & 1.38 & 1.36 & 0.38 & 0.47 & 0.15 & 1.68 \\ \hline
\textbf{Beacon} & \textit{E} & \textit{K} & \textit{G} & \textit{E} & \textit{K} & \textit{G} & \textit{E} & \textit{K} & \textit{G} & \textit{E} & \textit{K} & \textit{G} & \textit{E} & \textit{K} & \textit{G} & \textit{E} & \textit{K} & \textit{G} \\ \hline
\end{tabular}%
}
\label{env1all}
\end{table*}

\subsection{Discussion}
After the experiments in the two rooms, it is clear that all proximity measurements were improved substantially with filtering, specifically with the NI and PF. To get a better sense of the accuracy of all filters in both environments, two variations of the error were utilized. The MAE produces a smaller error in all cases. This may be because the MAE gives all instances of error overall proximity measurements equal weight. Since the RMSE squares the errors, especially large errors are given a substantially larger weight. Since this study is with regards to proximity and not position, MAE is sufficient in providing a good understanding of the true error, since the transmission intervals are so great, and some RSSI observations may produce significant outliers in comparison to the entire sequence leading up to the most recent.

In both environments, the Gimbal beacon achieved the smallest error under both error calculations. An important observation is that after 1.5~m the proximity estimation accuracy falls significantly. This suggests that the strategic placement of numerous beacons in a given environment should be considered for high accuracy LBS. However, the effects of beacon interference are to be expected and should be tested in order to determine the optimal layout and filter tuning parameters. Another underlying trend is that the Estimote beacon seems to perform significantly better in larger environments over smaller ones. This would require further tests to confirm, but may lead to further understanding of possible connections between room size and optimal transmission power. 

Although the Gimbal beacon is the cheapest of the three and lacks in features and sensors compared to the two alternatives, it consistently performs the best in these experiments. This may not be true for all environments, and so the requirement of environmental testing is deemed necessary for complex applications. However, the use of NI or particle filter will likely improve the performance of any beacon in many environments. There was no determinate winner but the NI and particle filter appeared to be the best performing overall. This was expected due to their non-parametric approximation of the system, which suits the path loss model well.

The inclusion of three beacon hardware devices in this experiment attempts to account for the variability in performance that may be experienced with other BLE beacon devices on the market. It can be seen that the proximity estimation accuracy increase when Bayesian filtering is used for all beacons tested and in both environments. This result may imply that alternate beacon devices should behave similarly, but it can not be determined without experimentation. It is important to note that all experiments were conducted in a semi-controlled environment. Alternative deployment scenarios may present a greater amount of system noise and physical obstructions, thus reducing accuracy. Hence the need for predictive filters.

\begin{table}[t!]
\centering%
\scriptsize%
\renewcommand{\arraystretch}{1}%
\setlength{\tabcolsep}{0.2\tabcolsep}%
\settowidth{\cellwidtha}{Display resolution (pixels)}%
\setlength{\cellwidthb}{\columnwidth}%
\addtolength{\cellwidthb}{-\cellwidtha}%
\addtolength{\cellwidthb}{-2\tabcolsep}%
\setlength{\cellwidthb}{0.23\cellwidthb}%
\addtolength{\cellwidthb}{-2\tabcolsep}%
\caption{Environment 2: MAE (\textnormal{m}).}%
\begin{tabular}{|l|c|c|c|c|c|c|} \hline
\makebox[\cellwidthb][l] & \makebox[\cellwidthb][c]{SMA} & \makebox[\cellwidthb][c]{KF-ST} & \makebox[\cellwidthb][c]{KF-DN}& \makebox[\cellwidthb][c]{NIF}& \makebox[\cellwidthb][c]{PF}\\ \hline \hline
Estimote & 0.842 & 0.544 & 0.543 & 0.548 & \textbf{0.529}\\ \hline
Kontakt  & 0.579 & 0.405 & 0.402 & \textbf{0.379} & 0.391\\ \hline
Gimbal   & 0.359 & 0.335 & 0.343 & 0.317 & \textbf{0.315}\\ \hline \hline
\textit{Average}   & \textit{0.593} & \textit{0.428} & \textit{0.429} & \textit{0.415} & \textit{\textbf{0.412}}\\ \hline 

\end{tabular}%
\label{meanTable_r2}%
\end{table}

\begin{table}[t!]
\centering%
\scriptsize%
\renewcommand{\arraystretch}{1}%
\setlength{\tabcolsep}{0.2\tabcolsep}%
\settowidth{\cellwidtha}{Display resolution (pixels)}%
\setlength{\cellwidthb}{\columnwidth}%
\addtolength{\cellwidthb}{-\cellwidtha}%
\addtolength{\cellwidthb}{-2\tabcolsep}%
\setlength{\cellwidthb}{0.23\cellwidthb}%
\addtolength{\cellwidthb}{-2\tabcolsep}%
\caption{Environment 2: RMSE (\textnormal{m}).}%
\begin{tabular}{|l|c|c|c|c|c|c|} \hline
\makebox[\cellwidthb][l] & \makebox[\cellwidthb][c]{SMA} & \makebox[\cellwidthb][c]{KF-ST} & \makebox[\cellwidthb][c]{KF-DN}& \makebox[\cellwidthb][c]{NIF} & \makebox[\cellwidthb][c]{PF}\\ \hline \hline
Estimote & 1.407 & 0.869 & 0.873 & \textbf{0.797} & 0.807\\ \hline
Kontakt  & 0.765 & 0.584 & 0.583 & \textbf{0.554} & 0.627\\ \hline
Gimbal   & 0.594 & 0.549 & 0.598 & 0.539 & \textbf{0.517}\\ \hline \hline
\textit{Average}   & \textit{0.922} & \textit{0.667} & \textit{0.685} & \textit{\textbf{0.63}} & \textit{0.65}\\ \hline 

\end{tabular}%
\label{rmseTable_r2}%
\end{table}

\section{Conclusion}\label{concl} 
In this work, three Bayesian filtering techniques were implemented locally on a mobile device to improve proximity estimation using RSSI techniques.  Three different beacons were tested  in two environments, a small and a large room. The use of particle filtering and non-parametric filtering had a significant improvement in proximity estimation. The experimental results enforced the necessity of a well-developed path loss model for RSSI-based distance estimation. Overall, proximity accuracy can be vastly improved with filtering techniques. Experimental results depicted an improvement of up to 40\% over the smoothed results, achieving proximity error range as low as 0.27~m when the beacon is within 3~m from the receiver. BLE beacons are a promising solution for PBS since they are cheap and easy to deploy, with low energy requirements. The use of Bayesian filtering can improve their performance, without significant overhead.

\bibliographystyle{IEEEtran}
\bibliography{IEEEabrv,sensorsbib}

\end{document}